\documentclass{article}
\usepackage{spconf,amsmath,graphicx}
\usepackage{multirow}
\usepackage{booktabs}
\usepackage{xcolor}
\usepackage{textcomp}
\usepackage{algorithmic}
\usepackage{bm}
\usepackage{soul}
\usepackage{cite}
\usepackage{amssymb,amsfonts}
\usepackage{caption}
\usepackage{subcaption}
\usepackage[switch,pagewise]{lineno}
\usepackage{algorithm}
\usepackage{threeparttable}

\title{DCNGAN: A DEFORMABLE CONVOLUTION-BASED GAN WITH QP ADAPTATION FOR  PERCEPTUAL QUALITY ENHANCEMENT OF COMPRESSED VIDEO}

\name{Saiping Zhang$^{1}$, Luis Herranz$^{2}$ \thanks{L.H. acknowledges the support of grants RYC2019-027020-I and RTI2018-102285-A-I00 (MICINN, Spain).}, Marta Mrak$^{3}$, Marc G\'orriz Blanch$^{3}$, Shuai Wan$^{4}$, Fuzheng Yang$^{1}$}
\address{$^{1}$State Key Laboratory of Integrated Services Networks, Xidian University, Xi’an, China\\$^{2}$Computer Vision Center, Universitat Autònoma de Barcelona, 08193 Barcelona, Spain\\$^{3}$BBC Research \& Development, The Lighthouse, White City Place, 201 Wood Lane, London, UK\\$^{4}$School of Electronics and Information, Northwestern Polytechnical University, Xi’an, China}

\begin{document}
%
\maketitle
\begin{abstract}
In this paper, we propose a deformable convolution-based generative adversarial network (DCNGAN) for perceptual quality enhancement of compressed videos. DCNGAN is also adaptive to the quantization parameters (QPs). Compared with optical flows, deformable convolutions are more effective and efficient to align frames. Deformable convolutions can operate on multiple frames, thus leveraging more temporal information, which is beneficial for enhancing the perceptual quality of compressed videos. Instead of aligning frames in a pairwise manner, the deformable convolution can process multiple frames simultaneously, which leads to lower computational complexity. Experimental results demonstrate that the proposed DCNGAN outperforms other state-of-the-art compressed video quality enhancement algorithms.
\end{abstract}
\section{Introduction}
Recent years have witnessed the tremendous development of video compression algorithms \cite{b1}\cite{b2}. However, compressed videos, especially at low bit rate, still suffer from the degraded quality due to compression artifacts. In this case, it is crucial to enhance the quality of compressed videos.

Previous works focus on enhancing the objective quality of compressed videos \cite{b3}\cite{b4}\cite{b5}. Yang et al. \cite{b3} proposed a compressed video quality enhancement algorithm which aggregated information from neighboring high quality frames, named MFQE. By further optimizing the network in \cite{b3}, Guan et al. \cite{b4} proposed MFQE 2.0 and achieved better performance. Deng et al. \cite{b5} incorporated deformable convolutions \cite{b12} to efficiently enhance the PSNR of compressed videos. However, sometimes the objective quality is inconsistent with the perceptual quality \cite{b6}. To achieve higher quality of experience (QoE) \cite{b21}, many works have aimed at enhancing the perceptual quality of compressed videos \cite{b7}\cite{b8}. Wang et al. \cite{b7} proposed a generative adversarial network (GAN) based on the multi-level wavelet packet transform to enhance compressed videos. Wang et al. \cite{b8} removed visual artifacts by an enhancement network with residual blocks.

However, those algorithms require training and storing various models to enhance videos compressed at different QPs, which sets high demand on the memory. Adapting to specific QPs, Huang et al. \cite{b10} achieved QP adaptive CNN-filters by feeding the coded QP to the network. Liu et al. \cite{b11} embedded the quantization step (Qstep) into the network to adapt QP and improve coding performance.

\begin{figure*}[]
\centering
\includegraphics[width=1\textwidth]{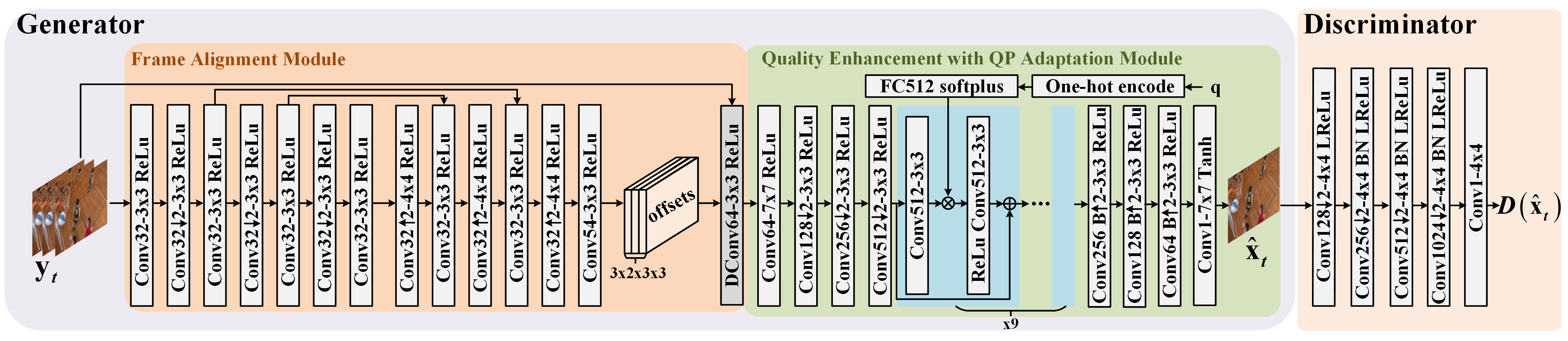}
\caption{The proposed DCNGAN. Conv\textit{N}-$K \times K$ is a convolution with \textit{N} channels, $K \times K$ filters. ↓2 and ↑2 are strided down and up by a factor 2, respectively. DConv is the deformable convolution. FC\textit{N} is the fully connected layer with \textit{N} dimensions. B↑2 is Bilinear upsampling by a factor 2. LReLu is Leaky ReLu with $\alpha=0.2$. BN is Batch Normalization.}
\label{fig 1}
\end{figure*}

In this paper, we propose a QP-adaptive deformable convolution-based GAN to enhance the perceptual quality of compressed videos. Thanks to QP adaptation, our method can  enhance videos displayed on TV, where QP values vary greatly from frame to frame. Moreover, our method can be used for historical content enhancement to meet higher quality requirements. Our main contributions are:
\begin{enumerate}
\item A GAN framework based on deformable convolutions to enhance the perceptual quality of compressed videos.

\item A single adaptive model to enhance videos compressed at various QPs.

\item We compare the proposed DCNGAN with state-of-the-art compressed video quality enhancement networks, showing its superior performance.
\end{enumerate}
\section{Proposed framework}

Our framework is illustrated in Fig. 1, and consists of a generator $\hat{\mathbf{x}}_t=G\left(\mathbf{y}_t,\mathbf{q}\right)$ and a discriminator $D\left(\hat{\mathbf{x}}_t\right)$. At a particular frame index $t$, the generator processes a sequence of three consecutive frames $\mathbf{y}_t=\left(\mathbf{x}_{t - 1},\mathbf{x}_{t},\mathbf{x}_{t + 1}\right)$ and the corresponding QP value $\mathbf{q}$ to output the enhanced frame $\hat{\mathbf{x}}_t$. 

\subsection{Generator}

The generator $G\left(\mathbf{y}_t\right)$ further consists of two modules: the frame alignment module $\mathbf{z}_t=A\left(\mathbf{y}_t\right)$ and the quality enhancement module $\hat{\mathbf{x}}_t=E\left(\mathbf{z}_t\right)$. In order to leverage temporal information, the former takes previous, current and next frames and aligns them using a deformable convolution. The output of the deformable convolution is a single representation $\mathbf{z}_t$ which already integrates information from the three frames. The key to the correct alignment is the computation of the corresponding offsets, which are predicted by a network based on U-net \cite{b13}. The number of channels of offsets is determined by the number of the input frames (i.e. three in our case), the number of spatial dimensions (i.e. two) and the size of the convolutional kernels (i.e. $3\times 3$ in our case). This approach is more efficient than aligning them using optical flows (generally conducted in a pairwise manner).

The quality enhancement module is based on an encoder-decoder structure with nine residual blocks. To avoid storing multiple sets of model parameters for enhancing videos compressed at different QPs, encoded QP information $\mathbf{q}$ is embedded into each residual block to make the network modulated by QP values. Specifically, $\mathbf{q}$ is represented using one-hot encoding \cite{b10} and fed into a fully connected layer. Softplus is selected as the activation function since it ensures positive outputs which we found beneficial in training. Finally, channel-wise multiplication is performed between the feature maps which are the output of the first convolution layer before ReLU in the residual block and the encoded QP information.

\subsection{Discriminator}
We use a patch discriminator \cite{b14}  which outputs the probability of each patch being real or fake. It is implemented in a fully convolutional fashion, and the final output is the average real/fake classification probability across patches.

\subsection{Objective}
Our model optimizes the following adversarial loss \cite{b15} (hereinafter we omit the subindex $t$ for simplicity)
\begin{equation}
{L_{gan}}\left( {G,D} \right) = {\rm{\mathbb{E}}}_{\left({\mathbf{y},\mathbf{q}}\right)}\left[ {{{\left( {D\left( {G\left( \mathbf{y},\mathbf{q} \right)} \right) - 1} \right)}^2}} \right] + {\rm{\mathbb{E}}}_{\mathbf{x}}\left[ {D{{\left( \mathbf{x} \right)}^2}} \right],
\end{equation}
where $\mathbf{y}$ are sequences of three consecutive decoded frames and $\mathbf{q}$ is the corresponding QP value, and $\mathbf{x}$ are raw frames, both obtained from a video dataset. The generator is trained to minimize the value of Eq. (1) while the discriminator is trained to maximize it.


We use a perceptual loss based on VGG features \cite{b16} that enforces that the features of a given pair $\left(\mathbf{y},\mathbf{q}\right)$ match the features of the corresponding target raw frame $\mathbf{x}$. 
\begin{equation}
{L_{vgg}}\left( G \right) = {\rm{\mathbb{E}}_{\left( {{\bf{y}},{\bf{q}},{\bf{x}}} \right)}}\sum\limits_{i{\rm{ = 1}}}^{{N_f}} {\frac{{\rm{1}}}{{{M_i}}}\sum\limits_{j{\rm{ = 1}}}^{{M_i}} {{{\left\| {f_j^i\left( {\bf{x}} \right){\rm{ - }}f_j^i\left( {G\left( {{\bf{y}},{\bf{q}}} \right)} \right)} \right\|}_{\rm{1}}}}},
\end{equation}
where ${f^i_j}\left(\cdot  \right)$ represents the $j$-th spatial element of the output tensor of the $i$-th layer from a pre-trained VGG-19 model.

Similarly, we also enforce  matching the features of enhanced and of raw images in the discriminator.
\begin{equation}
{L_{fm}}\left( {G,D} \right) = {\rm{\mathbb{E}}_{\left( {{\bf{y}},{\bf{q}},{\bf{x}}} \right)}}\sum\limits_{i{\rm{ = 1}}}^{{N_g}} {\frac{{\rm{1}}}{{M_i^g}}\sum\limits_{j{\rm{ = 1}}}^{M_i^g} {{{\left\| {g_j^i\left( {\bf{x}} \right){\rm{ - }}g_j^i\left( {G\left( {{\bf{y}},{\bf{q}}} \right)} \right)} \right\|}_{\rm{1}}}} },
\end{equation}
where ${g^i_j}\left(\cdot  \right)$ represents the $j$-th spatial element of the output tensor of the $i$-th layer selected from the discriminator.


Finally, during training we optimize
\begin{equation}
{\underset {G}{\operatorname {min} }}\ {\underset {D}{\operatorname {max}}}\ \left({L_{gan}}\left( {G,D} \right) + {L_{vgg}}\left( G \right) + {L_{fm}}\left( {G,D} \right)\right).
\end{equation}

\setlength\tabcolsep{2.0pt}
\begin{table*}[t]
\caption{Overall performance on LPIPS and DISTS of JCT-VC standard test sequences at four QPs}
\begin{center}
\begin{threeparttable}[b]
\begin{tabular}{ccccccccccccccc}
\toprule
\multirow{2}{*}{QP} & \multicolumn{2}{c}{\multirow{2}{*}{Sequences}} & \multicolumn{2}{c}{Compressed}  & \multicolumn{2}{c}{MFQE 2.0 \cite{b4}} & \multicolumn{2}{c}{STDF \cite{b5}} & \multicolumn{2}{c}{MW-GAN \cite{b7}} & \multicolumn{2}{c}{VPE-GAN \cite{b8}} & \multicolumn{2}{c}{Proposed} \\
\multirow{20}{*}{32} & \multicolumn{2}{c}{}                           & LPIPS         & DISTS    & LPIPS         & DISTS    & LPIPS       & DISTS      & LPIPS        & DISTS       & LPIPS        & DISTS        & LPIPS        & DISTS       \\
\midrule
                     & \textit{Class A}                    & \textit{Traffic}   &0.170       &0.014       &0.184               &0.014              &0.094             &0.009            &0.138              &—             &0.179              &0.029                          &\textbf{0.070}              &\textbf{0.006}              \\
                     &                            & \textit{PeopleOnStreet}     &0.150       &0.018       &0.167               &0.018              &0.133             &0.010            &0.130              &—             &0.135              &0.015                         &\textbf{0.086}              &\textbf{0.008}            \\
                     & \multirow{5}{*}{\textit{Class B}}   & \textit{Kimono}    &0.258       &0.043       &0.294               &0.046              &0.160             &0.026            &0.189              &—             &0.180              &0.034                          &\textbf{0.108}              &\textbf{0.023}              \\
                     &                            & \textit{ParkScene}          &0.276       &0.044       &0.286               &0.045              &0.182             &0.027            &0.244              &—             &0.196              &0.037               &\textbf{0.123}            &\textbf{0.023}              \\
                     &                            & \textit{Cactus}             &0.260       &0.022       &0.288               &0.022              &0.136             &0.012            &0.151              &—             &0.126              &0.017                    &\textbf{0.096}            &\textbf{0.010}              \\
                     &                            & \textit{BQTerrace}          &0.215       &0.032       &0.241               &0.034              &0.152             &0.021            &0.116              &—             &0.140              &0.040                        &\textbf{0.113}          &\textbf{0.018}            \\
                     &                            & \textit{BasketballDrive}    &0.247       &0.028       &0.279               &0.031              &0.166             &0.022            &0.141              &—             &0.132              &0.025                     &\textbf{0.099}          &\textbf{0.015}             \\
                     & \multirow{4}{*}{\textit{Class C}}   & \textit{RaceHorses}&0.147       &0.066       &0.174               &0.075              &0.120             &0.061            &0.126              &—             &0.101              &0.055                &\textbf{0.089}               &\textbf{0.042}            \\
                     &                            & \textit{BQMall}             &0.124       &0.066       &0.145               &0.071              &0.089             &0.050            &0.091              &—             &0.112              &0.063                    &\textbf{0.072}              &\textbf{0.038}              \\
                     &                            & \textit{PartyScene}         &0.101       &0.057       &0.126               &0.060              &0.067             &0.042            &\textbf{0.026}   &—             &0.091              &0.045                         &0.075                     &\textbf{0.029}           \\
                     &                            & \textit{BasketballDrill}    &0.156       &0.073       &0.181               &0.079              &0.126             &0.068            &0.109              &—             &0.105              &0.060                         &\textbf{0.072}          &\textbf{0.040}             \\
                     & \multirow{4}{*}{\textit{Class D}}&\textit{\textit{RaceHorses}}&0.122  &0.121       &0.143               &0.132              &0.098             &0.113            &0.117              &—             &0.093              &0.126                     &\textbf{0.072}             &\textbf{0.091}              \\
                     &                            & \textit{BQSquare}           &0.110       &0.150       &0.121               &0.160              &0.084             &0.130            &0.073              &—             &\textbf{0.066}              &\textbf{0.112}               &0.104             &0.123            \\
                     &                            & \textit{BlowingBubbles}     &0.102       &0.117       &0.111               &0.128              &0.068             &0.104            &\textbf{0.063}              &—             &0.072              &0.096                    &0.065           &\textbf{0.084}            \\
                     &                            & \textit{BasketballPass}     &0.116       &0.135       &0.135               &0.150              &0.099             &0.127            &0.095              &—             &0.085              &0.116                     &\textbf{0.067}             &\textbf{0.099}           \\
                     & \multirow{3}{*}{\textit{Class E}}   & \textit{FourPeople} &0.120      &0.037       &0.128               &0.038              &0.089             &0.022            &0.080              &—             &0.103              &0.028              &\textbf{0.054}             &\textbf{0.016}            \\
                     &                            & \textit{Johnny}             &0.148       &0.035       &0.159               &0.035              &0.111             &0.021            &0.083              &—             &0.178              &0.059                           &\textbf{0.063}         &\textbf{0.014}             \\
                     &                            & \textit{KristenAndSara}    &0.134        &0.038       &0.148               &0.039              &0.106             &0.025            &0.108              &—             &0.136              &0.046                         &\textbf{0.062}          &\textbf{0.019}            \\
                     & \multicolumn{2}{c}{\textit{Average}}                    &0.164        &0.061       &0.184               &0.065              &0.116             &0.049            &0.115              &—             &0.124              &0.056                      &\textbf{0.083}           &\textbf{0.039}              \\
\hline
22                   & \multicolumn{2}{c}{\textit{Average}}                    &0.077        &0.020  &0.087                    &0.022                   &0.050        &\textbf{0.014}       &—                   &—             &0.097              &0.047                     &\textbf{0.042}        &0.017           \\

27                   & \multicolumn{2}{c}{\textit{Average}}                    &0.116        &0.037  &0.130                    &0.040                   &0.077        &0.029       &—                   &—             &0.103              &0.054                      &\textbf{0.059}       &\textbf{0.026}           \\
    
37                   & \multicolumn{2}{c}{\textit{Average}}                    &0.223        &0.089  &0.232               &0.086              &0.168        &0.080       &0.177              &—             &0.148              &0.070           &\textbf{0.120}           &\textbf{0.058}            \\
\bottomrule 
\end{tabular}
\end{threeparttable}
\end{center}
\end{table*}

\begin{figure*}[t]
\centering
\includegraphics[width=0.90\textwidth]{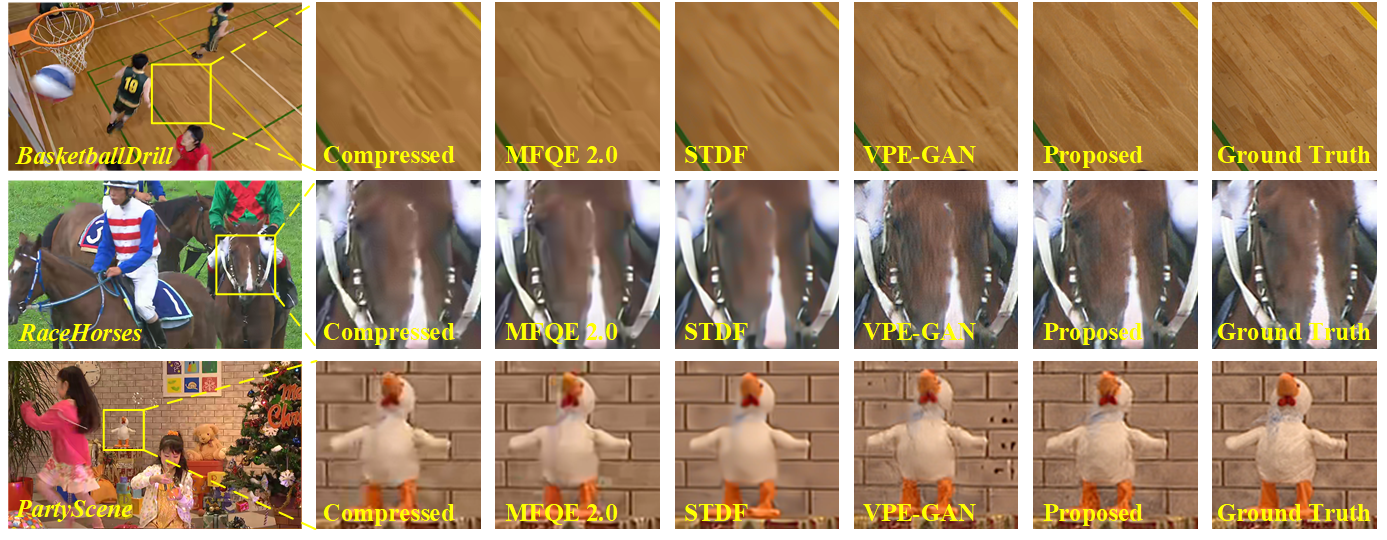}
\caption{Comparison results of the proposed DCNGAN with the other video enhancement algorithms at QP 37.}
\label{fig 2}
\end{figure*}

\section{Experimental Results}
\subsection{Dataset and Training Strategy}
We employ 106 sequences collected by \cite{b4} for training and the Joint Collaborative Team on Video Coding (JCT-VC) standard test sequences for testing. All sequences are compressed by H.265/HEVC software HM16.5 under Low Delay P configuration at QPs 22, 27, 32 and 37. Note that the 106 raw sequences and their compressed versions are randomly cropped into $128 \times 128$ clips as training samples. The training dataset includes mixed and shuffled samples at the four QPs.

We employ Adam optimizer with ${\beta _1} = 0.9$, ${\beta _2} = 0.999$ and ${\epsilon} = {10^{ - 8}}$. Batch size is set to be 32. Learning rate remains unchanged at ${10^{ - 4}}$. For a fair comparison with previous work, we only enhance the luminance component.

\subsection{Quantitative and Qualitative Comparison}
We compare the proposed DCNGAN with state-of-the-art video quality enhancement networks in \cite{b4} (MFQE 2.0), \cite{b5} (STDF), \cite{b7} (MW-GAN) $\footnote{For fair comparison, the performance of MW-GAN shown in Table 1 is from their published paper since our retrained model has not achieved the same good performance and their pretrained models at the four QPs have not been published.}$ and \cite{b8} (VPE-GAN). LPIPS \cite{b18} and DISTS \cite{b19} are employed to quantitatively evaluate the perceptual quality of enhanced videos. For better illustration, LPIPS and DISTS of ``Compressed" (i.e. the input videos) are also shown in Table 1. Smaller values indicate better perceptual quality. It should be noted that we only train one model for testing sequences compressed at four different QPs, while the other networks need to train four models since their proposed network cannot achieve QP adaptation. As shown in Table 1, the proposed DCNGAN achieves the advanced performance and alleviates memory requirements at the enhancement stage.

Besides, three examples are randomly selected to qualitatively illustrate the performance of the proposed DCNGAN compared with the other video quality enhancement networks. Specifically, as shown in Fig. 2, MFQE 2.0 and STDF, which were designed to improve the PSNR of compressed videos, still tend to produce blurred results and penalize the perceptual quality. As for VPE-GAN, it alleviates blurring to some extent, but sometimes generates artifacts (e.g. the artifacts on the horse and the wall in Fig. 2). These artifacts also result in degraded perceptual quality, while the proposed DCNGAN can generate more realistic high-frequency details to combat with blurring and greatly improve the perceptual quality of compressed videos.

\subsection{QP Adaptation Performance Evaluation}
To evaluate the ability of QP adaptation of the proposed DCNGAN, we separately train four models at the four QPs, represented by ``Trained\_QP22", ``Trained\_QP27", ``Trained\_QP32" and ``Trained\_QP37", and compare these four models with the single model trained to adapt to various QPs, represented by ``Trained\_4QPs". The performance comparison is shown in Fig. 3, where LPIPS is averaged on all sequences in Table 1.

Overall, the model trained at a certain QP achieves the best performance tested at that QP because it has fully learned the characteristics of videos compressed at the corresponding QP in the training process, while the model can hardly achieve satisfactory performance in enhancing the perceptual quality of videos compressed at other QPs due to different characteristics. By feeding encoded QP information $\footnote{HM 16.5 compresses videos with a small QP variation. We have tried to feed QP of each frame to the network, but haven’t see benefit of retraining with small delta QP. Hence, we only feed the QP of I frame to the network.}$ into the proposed DCNGAN, the model ``Trained\_4QPs" can be employed to enhance the perceptual quality of compressed videos at various QPs with advanced performance.

\begin{figure}[]
\centering
\includegraphics[width=0.40\textwidth]{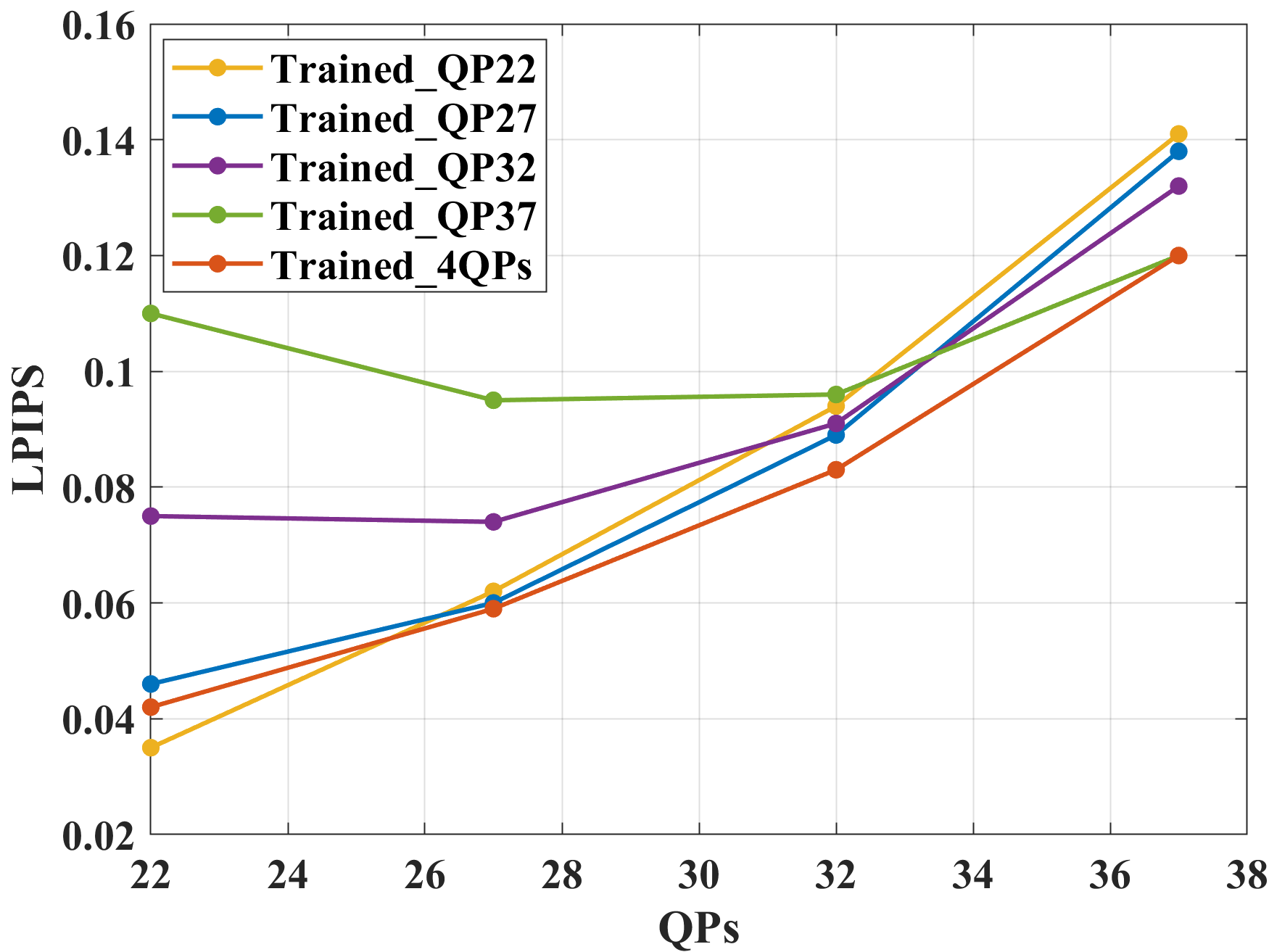}
\caption{LPIPS performance of models trained at different QPs.}
\label{fig 3}
\end{figure}

\begin{figure}[]
\centering
\includegraphics[width=0.40\textwidth]{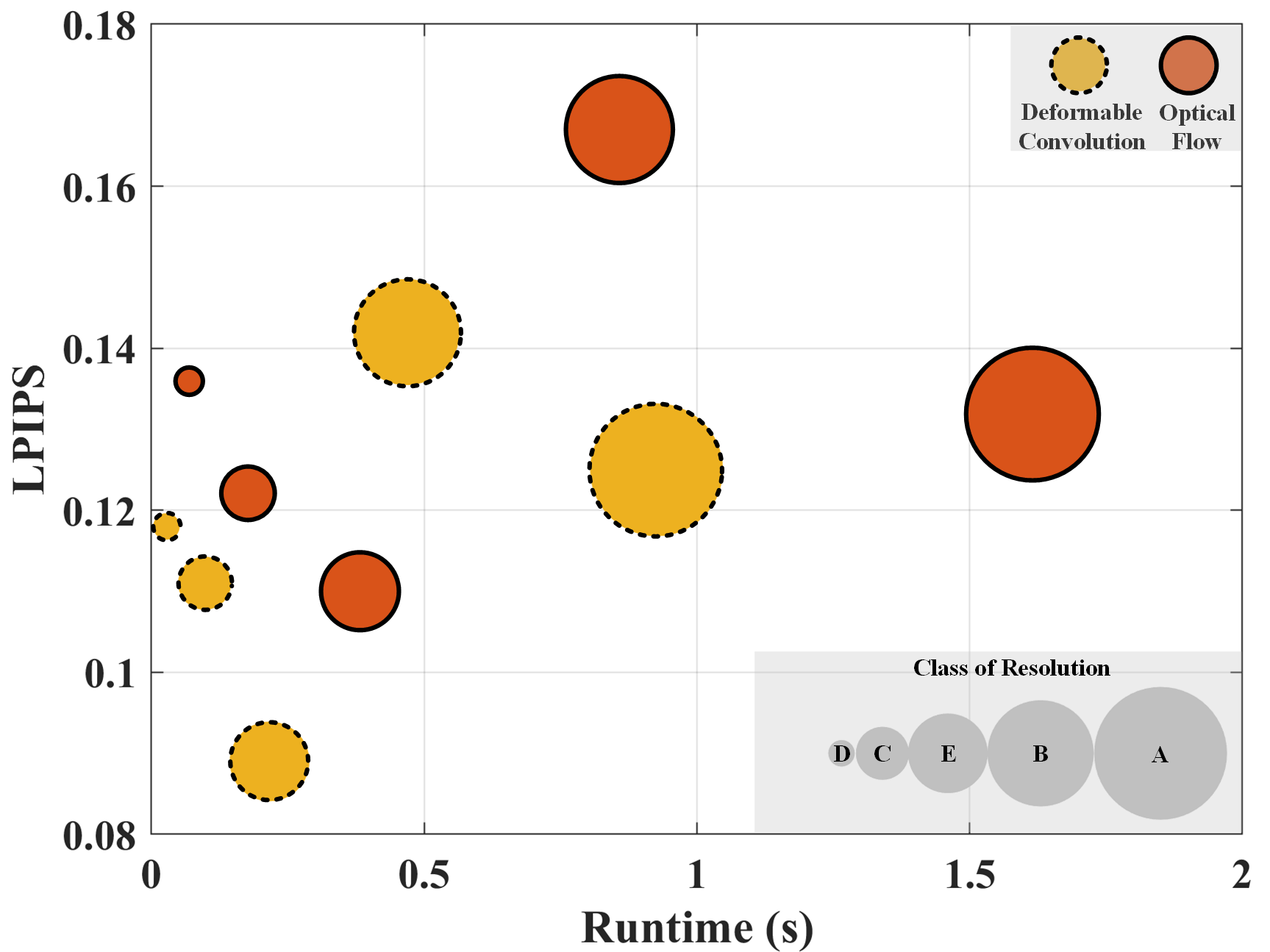}
\caption{LPIPS performance vs. runtime.}
\label{fig 4}
\end{figure}

\subsection{Deformable Convolutions vs. Optical Flows}

An alternative to deformable convolutions to align frame information is optical flow. For comparison, we replace deformable convolutions with optical flows estimated by a pre-trained SpyNet model \cite{b20}, in order to achieve frame alignments. The other modules of the proposed DCNGAN (i.e., the quality enhancement with QP adaptation module, the discriminator and the loss functions) are kept unchanged. Finally, the average LPIPS performance and runtime of two approaches are compared. As shown in Fig. 4, the network using deformable convolutions is much faster and achieves better performance than that using optical flows, which highlights the accuracy and efficiency of deformable convolutions.

\section{Conclusion}
In this paper, DCNGAN is proposed to enhance the perceptual quality of videos compressed at various QPs. By incorporating the deformable convolution to align temporal neighboring frames and feeding encoded QP to modulate the network, the proposed DCNGAN achieves the advanced performance and saves model parameters. Experimental results showed the superiority of the proposed DCNGAN compared with state-of-the-art compressed video quality enhancement networks.

\end{document}